\documentclass[aps,reprint,superscriptaddress]{revtex4-1}  

\usepackage{amsfonts,amssymb,amsmath}

\usepackage{amstext}
\usepackage{graphicx}

\usepackage{paralist}
\usepackage{gensymb}

\newcommand{\DM}{Dzyaloshinskii--Moriya}

\newcommand{\LLG}{Landau--Lifshitz--Gilbert}
\newcommand{\Bloch}{Bloch}
\newcommand{\Neel}{N\'{e}el}

\newcommand{\WKB}{Wentzel--Kramers--Brillouin}

\DeclareMathOperator{\sech}{sech}

\newcommand{\dd}{\; \mathrm{d}}
\newcommand{\ddsmall}{\, \mathrm{d}}
\newcommand{\ddnoskip}{\mathrm{d}}

\def\rcurs{{\mbox{$\resizebox{.09in}{.08in}{\includegraphics[trim= 1em 0 14em 0,clip]{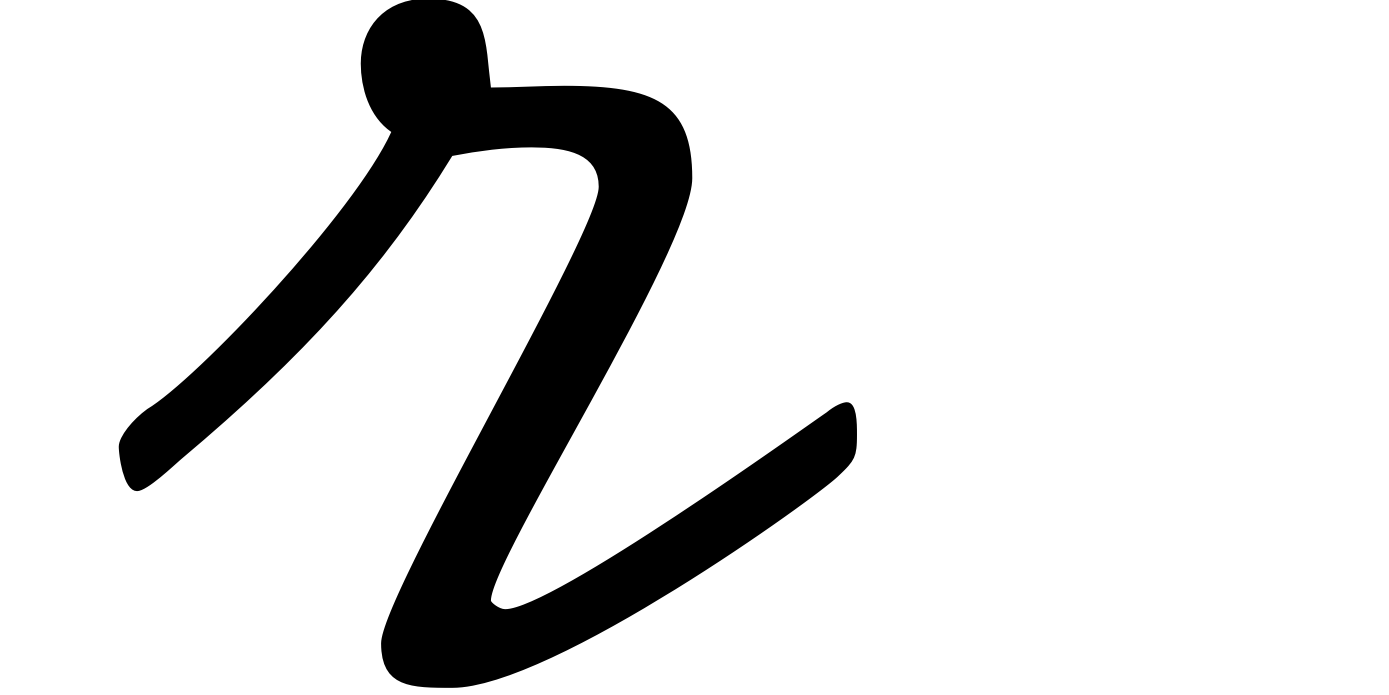}}$}}}
\def\brcurs{{\mbox{$\resizebox{.09in}{.08in}{\includegraphics[trim= 1em 0 14em 0,clip]{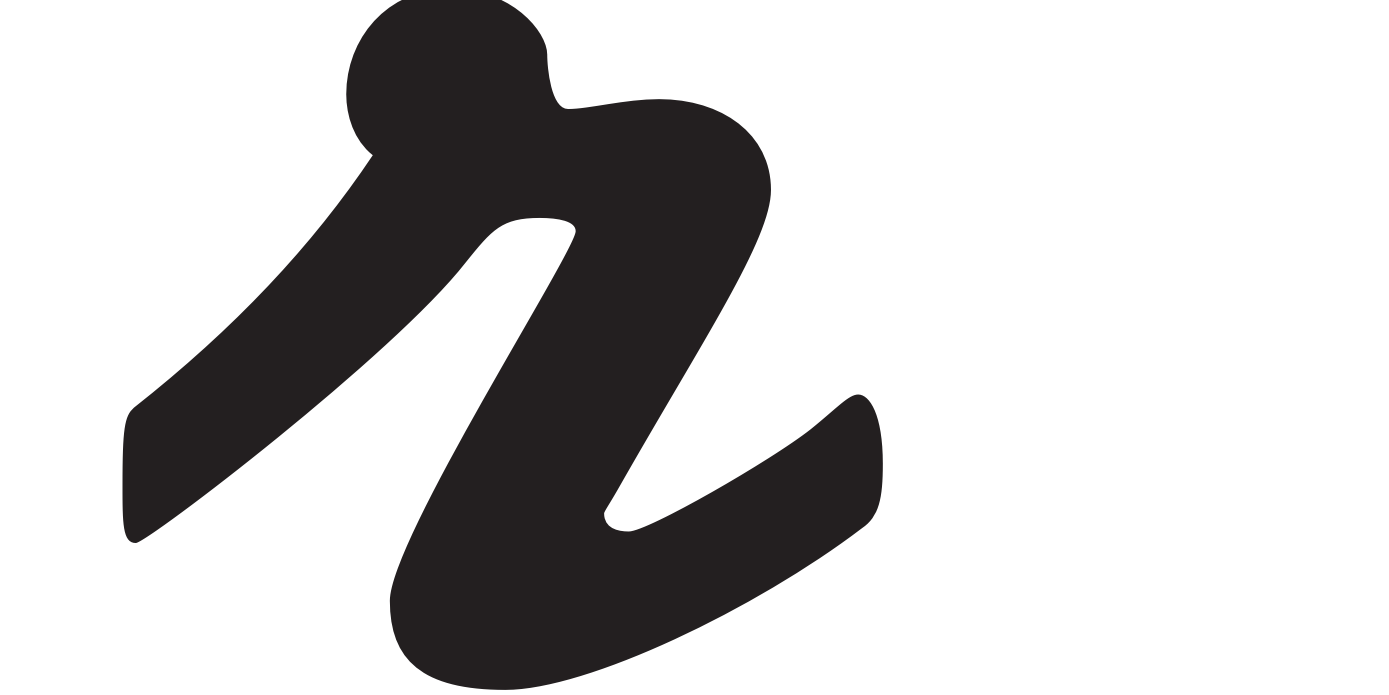}}$}}}

\newcommand{\secref}[1]{Sec.~\ref{#1}}

\newcommand{\equaref}[1]{Eq.~\eqref{#1}}
\newcommand{\Equaref}[1]{Equation~\eqref{#1}}

\newcommand{\explcite}[1]{Ref.~[\onlinecite{#1}]}

\newcommand{\explcites}[1]{Refs.~[\onlinecite{#1}]}

\newcommand{\explcitepart}[1]{[\onlinecite{#1}]}

\newcommand{\figref}[1]{Fig.~\ref{#1}}

\newcommand{\appendixheading}{\part*{Supplemental Material}
}

\begin{document}

\title{Chirality-dependent Transmission of Spin Waves through Domain Walls}

\author{F. J. Buijnsters}
\email[]{f.buijnsters@science.ru.nl}
\affiliation{Institute for Molecules and Materials, Radboud University, Heyendaalseweg~135, 6525~AJ Nijmegen, Netherlands}
\author{Y. Ferreiros}
\affiliation{Instituto de Ciencia de Materiales de Madrid, CSIC, Cantoblanco, 28049 Madrid, Spain}
\author{A. Fasolino}
\author{M. I. Katsnelson}
\affiliation{Institute for Molecules and Materials, Radboud University, Heyendaalseweg~135, 6525~AJ Nijmegen, Netherlands}

\date{March 7, 2016}


\begin{abstract}
Spin-wave technology (magnonics) has the potential to further reduce the size and energy consumption of information processing devices.
In the submicrometer regime (exchange spin waves), topological defects such as domain walls may constitute active elements to manipulate spin waves and perform logic operations.
We predict that spin waves that pass through a domain wall in an ultrathin perpendicular-anisotropy film experience a phase shift that depends on the orientation of the domain wall (chirality).
The effect, which is absent in bulk materials, originates from the interfacial \DM{} interaction and can be interpreted as a geometric phase.
We demonstrate analytically and by means of micromagnetic simulations that the phase shift is strong enough to switch between constructive and destructive interference.
The two chirality states of the domain wall may serve as a memory bit or spin-wave switch in magnonic devices.
\end{abstract}

\maketitle


Motivated by the aim to reduce energy dissipation in electronic devices, 
spin waves are considered as an alternative information carrier in the field of magnon spintronics~\cite{Chumak2015}.
A spin wave acquires a phase shift when it passes through a magnetic domain wall (DW) \cite{Hertel2004}.
In this Letter, we show that the \DM{} interaction (DMI) in ultrathin ferromagnetic films makes the phase shift dependent on the DW chirality, leading to constructive/destructive interference in a two-branch interferometer. 
The mechanism we identify raises the prospect of magnonic devices in which DW chirality acts as a spin-wave switch.

\begin{figure}
  \centering
  \includegraphics[scale=1.0]{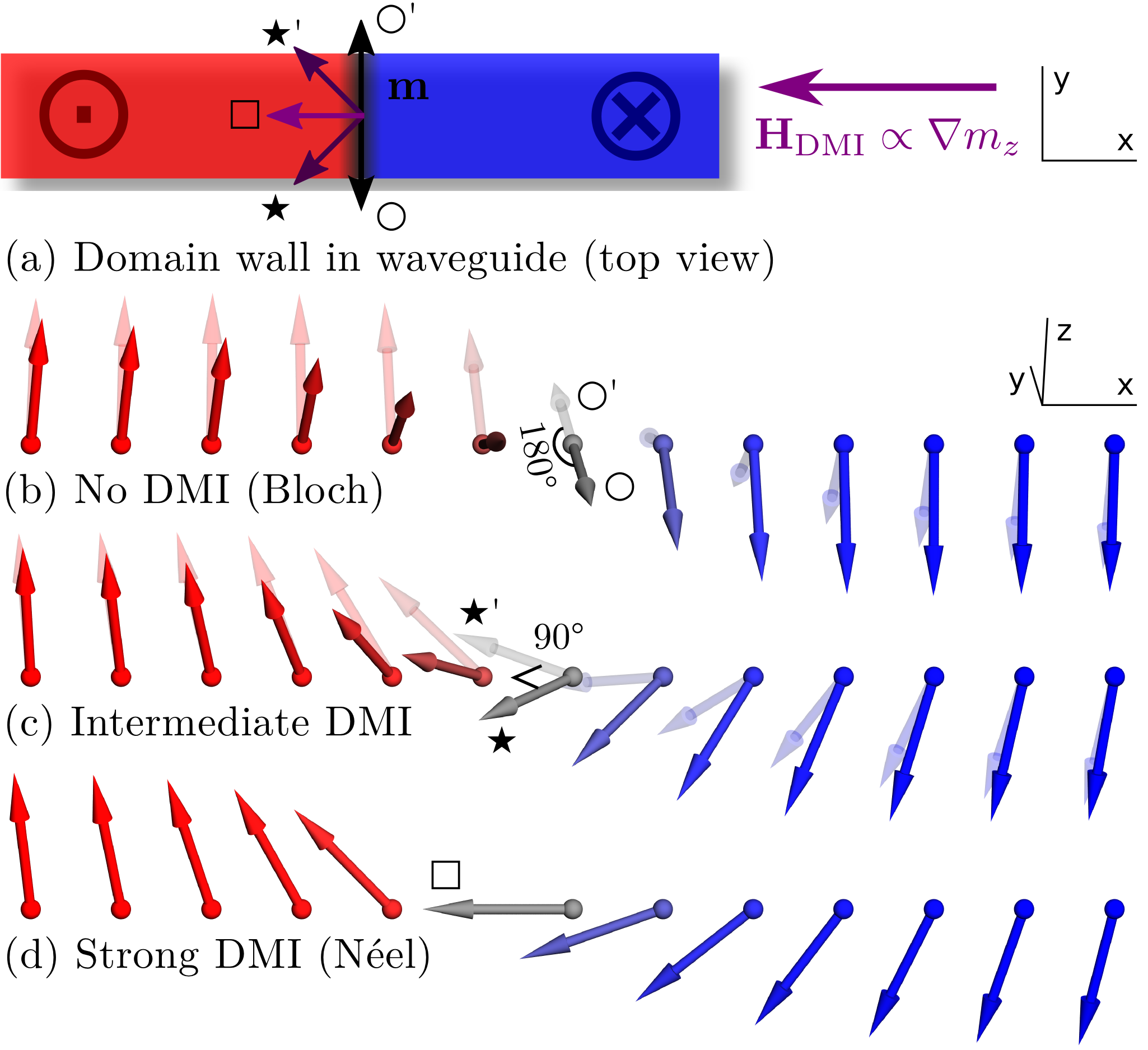}
  \caption{\label{fig:groundstate} (color online). 
  Effect of the interfacial DMI on the magnetization profile $\mathbf{m}(x)$ of a DW in a thin film with perpendicular anisotropy.
  (a)~Away from the DW, magnetization points out of the film ($\hat{\mathbf{z}}$ or $-\hat{\mathbf{z}}$).
  Near the DW, the DMI creates an effective field $\mathbf{H}_\text{DMI}$ in the $-\hat{\mathbf{x}}$ direction.
Depending on the competition between the dipolar and DMI interactions, the equilibrium configurations $\bigcirc,\bigcirc'$, $\bigstar,\bigstar'$, and $\square$, shown in (b)--(d), are possible.
  (b)~Without DMI, the minimum-energy configurations (flux closure) are two equivalent Bloch DWs ($\bigcirc$, in dark colors, and $\bigcirc'$, in light colors), whose in-plane orientations differ by $180^\circ$.
  (c)~For intermediate DMI, the minimum-energy configurations are intermediate between Bloch and \Neel{}.
  There are two equivalent minimum-energy states ($\bigstar$ and $\bigstar'$), whose in-plane orientations differ by approx.\ $90^\circ$ for an appropriately tuned DMI strength $D$.
  (d)~For strong DMI, a single minimum-energy configuration~$\square$ exists: a \Neel{} DW with magnetization in the center pointing in the $-\hat{\mathbf{x}}$ direction.
  }
\end{figure}
It is by now clear that the DMI plays a crucial role in
 the magnetization dynamics of ultrathin films \cite{Brataas2013,Emori2013,Thiaville2012,Heide2008}, due to the broken inversion symmetry at the interfaces.
The interfacial DMI favors, in a perpendicular-anisotropy film, \Neel{} DWs with a fixed chirality [Fig.~\ref{fig:groundstate}(d)] \cite{Thiaville2012,Emori2014}, in competition with the dipolar interaction, which tends to favor Bloch DWs [Fig.~\ref{fig:groundstate}(b)].
The most interesting regime is when the two interactions have a comparable strength, yielding a DW intermediate between Bloch and \Neel{} \cite{Heide2006,Emori2014} with two stable minimum-energy configurations (chirality states) whose in-plane orientations differ by  $\sim 90^\circ$, as shown in Fig.~\ref{fig:groundstate}(c).

Recent experiments demonstrated that DWs can be brought into the intermediate regime, and that the DMI strength can be fine-tuned by modifying the thicknesses of the adjacent nonmagnetic layers \cite{Franken2014}.
The internal orientation might be also tuned by an adjacent layer of a topological insulator; its surface states induce in the magnetic layer an interfacial-DMI-like effect that depends on chemical potential and applied electric field \cite{Tserkovnyak2012,Ferreiros2014,Linder2014,Ferreiros2015}.


\begin{figure}
  \centering
  \includegraphics[scale=1.0]{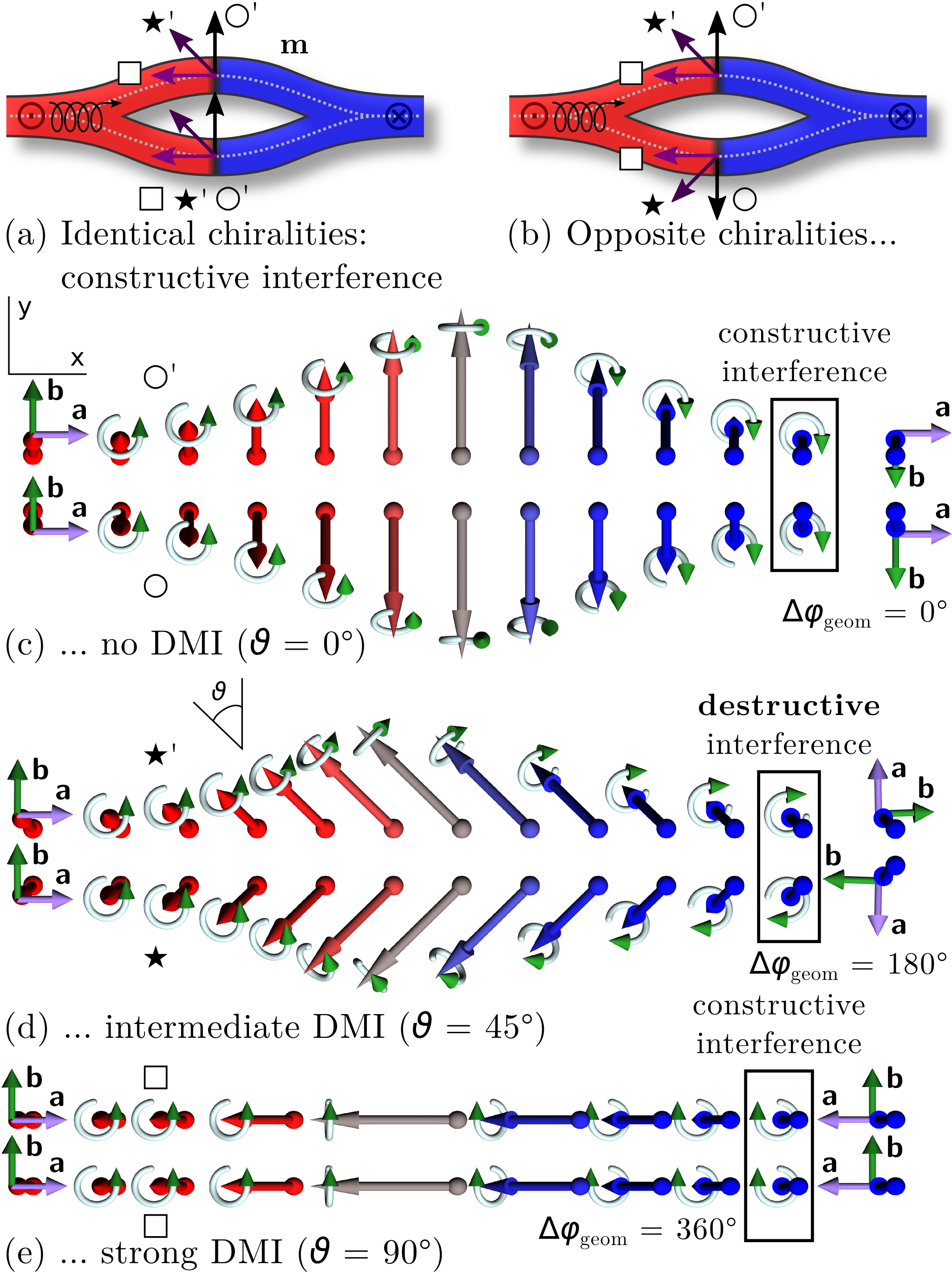}
  \caption{\label{fig:partrans} (color online).
  Interferometer setup in a thin film with perpendicular anisotropy. The two DWs  may have (a)~identical or (b)~opposite chiralities.
  Spin waves enter the device from the left.
  If the chiralities are identical, constructive interference is always obtained on the right-hand side.
  For opposite chiralities, the phase difference depends on the DMI strength, as shown in (c)--(e).
  (c)~Without DMI, spin waves interfere constructively even if the chiralities are opposite ($\bigcirc, \bigcirc'$).
  (d)~For intermediate DMI, 
  we find a phase difference $\Delta \varphi$ of up to $180^\circ$ (destructive interference) for opposite chirality states $\bigstar,\bigstar'$.
  (e)~For strong DMI, the configurations in both branches are the same ($\square$), trivially resulting in constructive interference.
  In (c)--(e), large arrows represent the equilibrium magnetization direction $\mathbf{m}(x)$.
  On the left, spin-wave basis vectors $\hat{\mathbf{a}},\hat{\mathbf{b}}$ are defined identically for all configurations $\bigcirc,\bigcirc',\bigstar,\bigstar',\square$.
  Their orientation after parallel transportation, shown on the right, depends on the DW configuration.
  In (d), notice that the transported basis vectors for $\bigstar$ and $\bigstar'$ are rotated by $180^\circ$.
  This geometric phase difference $\Delta \varphi_\text{geom}$ is the dominant contribution to $\Delta \varphi$.
  }
\end{figure}
Our main result is summarized in Fig.~\ref{fig:partrans}, where we consider an interferometer in which incoming spin waves are divided between two identical waveguides, each containing a DW. The two DWs are identical in every respect except possibly their chirality.
When the spin waves rejoin, they are transmitted or reflected depending on the phase difference.
While it is obvious that spin waves interfere constructively if the two DWs have the same chirality [Fig.~\ref{fig:partrans}(a)], we ask if it is possible to achieve destructive interference (spin wave blocked) by reversing the chirality of one DW [Fig.~\ref{fig:partrans}(b)].
Without DMI, the two chirality states of the Bloch DW induce identical phase shifts, leading to constructive interference [Fig.~\ref{fig:partrans}(c)].
For strong DMI, the DW has  a single stable (\Neel{}) configuration and the phase shifts are obviously also identical [Fig.~\ref{fig:partrans}(e)].
However, for intermediate DMI, the two equilibrium orientations induce geometric phase shifts differing by as much as $180^\circ$ [Fig.~\ref{fig:partrans}(d)].
In this regime, the interferometer can switch between constructive and destructive interference -- transmission or reflection -- depending on whether the chiralities are identical or opposite.

\begin{figure}
  \centering
  \includegraphics[scale=1.0]{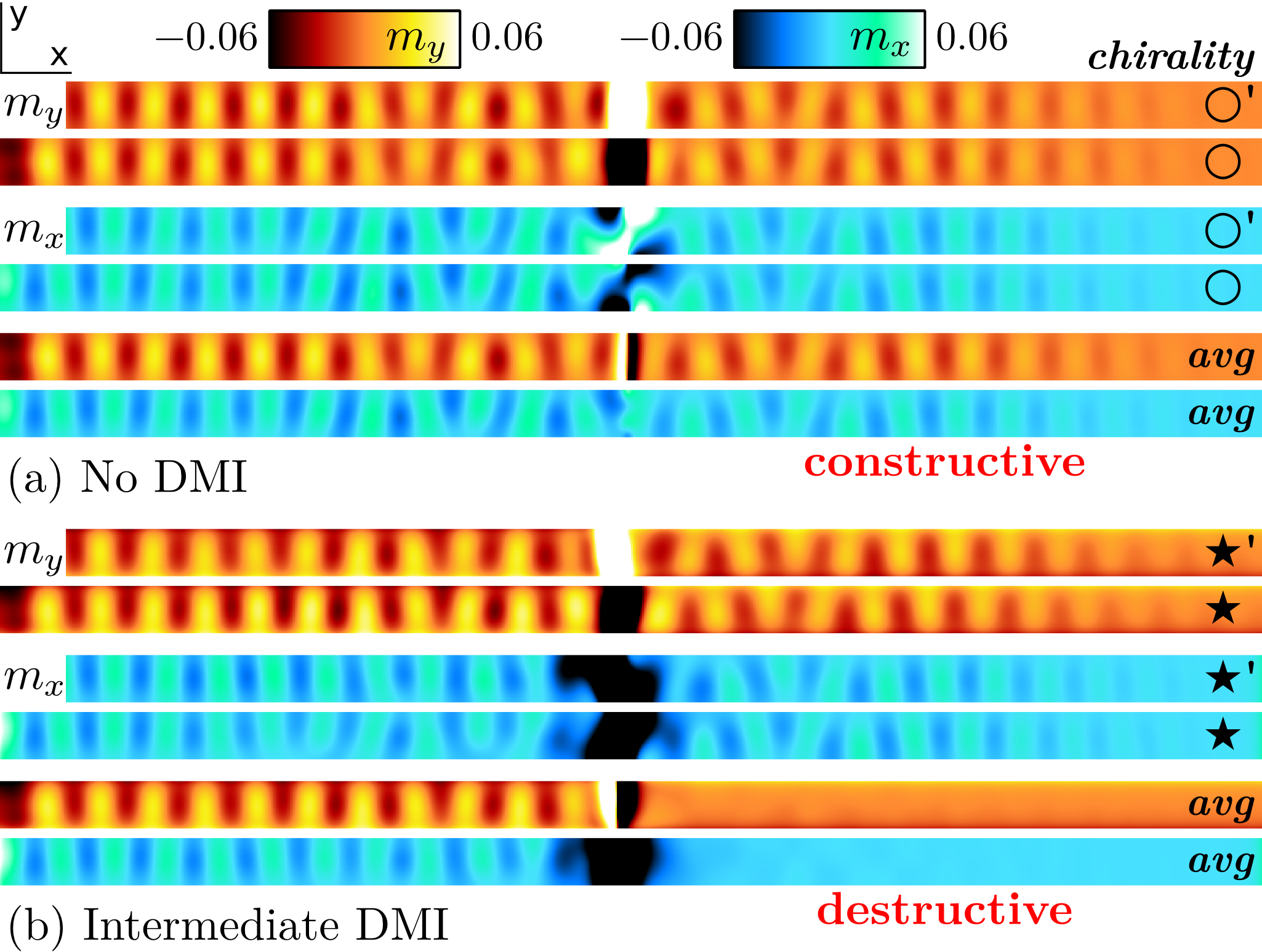}
  \caption{\label{fig:sims} (color online).
  Micromagnetic simulations of the propagation of spin waves in a ferromagnetic thin-film waveguide with perpendicular magnetization ($2\pi M_\text{S}^2 = 0.907\,K$, $L=3\,l$, where $l = \sqrt{A/K}$)  through DWs of the indicated chiralities ($\bigcirc',\bigstar'$ \emph{vs.}\  $\bigcirc,\bigstar$).
  (a)~Without DMI ($D=0$), spin waves experience the same phase shift regardless of DW chirality, leading to constructive interference (avg = average).
  (b)~For intermediate DMI ($D=0.06\,A/l$), there is a phase difference of almost $180^\circ$ between spin waves that passed through DWs of different chirality, leading to destructive interference.
  We remark that the attenuation on the right-hand side is not the result of Gilbert damping (we take $\alpha = 0.0030$), but merely represents the present location of the wavefront [$t = 89.6\,M_\text{S}/(|\gamma|K)$].
  }
\end{figure}

The two chirality states are separated by an energy barrier $\Delta E$ (an unfavorable \Neel{} configuration). 
If  $\Delta E$ is high enough, spontaneous reversals of chirality due to thermal fluctuations are very rare (for the system in Fig.~\ref{fig:sims}, we obtain $\Delta E 
= 2.5\times10^{-12}\text{ erg} = 61\, k_\text{B}T_\text{room}$).
We could consider the intermediate-DMI interferometer as a two-state memory device where the transmission of spin waves serves as readout mechanism (`open' or `closed').

\begin{figure*}
  \centering
  \includegraphics[scale=1.0]{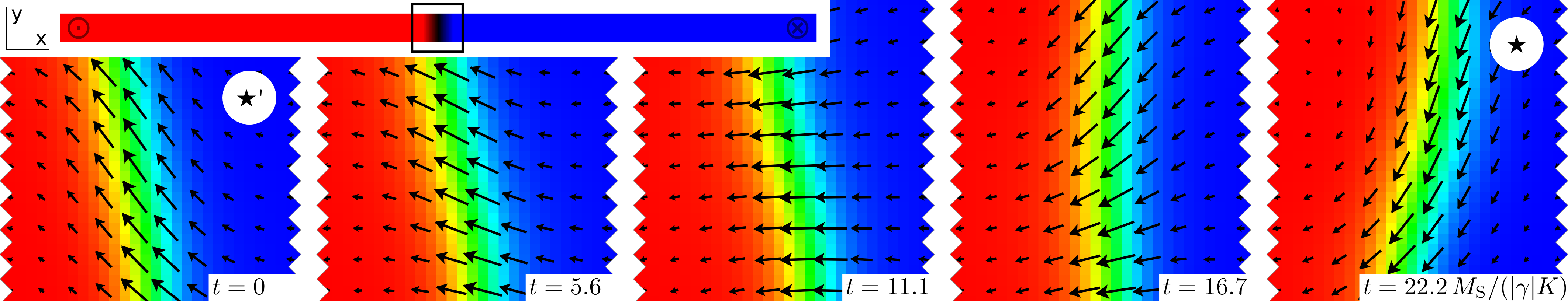}
  \caption{\label{fig:precess} (color online).
  Evolution of the magnetization during switching of DW chirality by means of an applied field $H_z$ perpendicular to the film in a two-dimensional micromagnetic simulation.
  Only a part of the waveguide is shown.
  The arrows represent the direction of the in-plane component of magnetization $\mathbf{m}(t,x,y)$ and the color the $z$ component.
  We take $H_\text{z} = 0.09 \, K/M_\text{S}$, $\alpha=0.2$; other parameters as in Fig.~\ref{fig:sims}(b).
  }
\end{figure*}

Switching does not require modifications of the material parameters,
nor to insert or remove DWs~\cite{Hertel2004}, but only to reverse the chirality of one DW, for instance 
by a field pulse normal to the plane of the film.
(The `field pulse' might alternatively be generated through optomagnetic effects \cite{Kirilyuk2010}, provided the light can be focused onto a single branch.)
The field causes the DW magnetization to precess as shown in Fig.~\ref{fig:precess} until, when it is switched off, 
the DW relaxes to the nearest chirality state.


We have tested the results of Fig.~\ref{fig:partrans}, which we derive analytically below, by means of explicit micromagnetic simulations.
The total energy $E$ is given by the sum of the usual micromagnetic energy functionals for exchange $E_\text{ex} = A\iint ({\lVert \partial_x \mathbf{m} \rVert}^2 + {\lVert \partial_y \mathbf{m} \rVert}^2) \ddsmall{}x \ddsmall{}y$, uniaxial anisotropy $E_\text{ani} = -K\iint m_z^2 \ddsmall{}x \ddsmall{}y$, and dipolar energy~\cite{supmat}, 
plus a functional
\begin{equation}\label{eq:interfacialDMI}
E_\text{DMI} = -2D \iint \mathbf{m} \cdot (\nabla m_z) \dd{}x \dd{}y 
\end{equation}
describing the DMI induced near the interfaces of the ultrathin film \cite{Thiaville2012}.
The DMI strength $D$ can be positive or negative (we take $D>0$ in Figs.~\ref{fig:groundstate} and~\ref{fig:partrans}).
We treat the film as effectively two-dimensional (magnetization is a function of $x$ and $y$ only), but we do consider the finite film thickness $L$  in the $z$ direction for the dipolar interactions \cite{supmat}.
Here we consider a waveguide made of a long strip of ultrathin film with perpendicular magnetization ($K > 2\pi M_\text{S}^2$, where $M_\text{S}$ is saturation magnetization).
The waveguide width $W$ is at least so large that the dipolar interactions, in the absence of DMI, favor a Bloch DW.

The interfacial DMI is qualitatively different from a DMI $\propto \iiint \mathbf{m} \cdot (\nabla \times \mathbf{m}) \ddsmall{}V$ present in isotropic bulk materials with a chiral crystal structure \cite{supmat}.
The effect of a bulk DMI on the interaction of spin waves with DWs was considered in \explcites{Yan2011} and~\explcitepart{Wang2015}.
Since a bulk DMI favors the Bloch DW ($\bigcirc$ in Figs.~\ref{fig:groundstate} and~\ref{fig:partrans}), it does not, in the geometry considered here, provide the
competition with dipolar interactions that is essential to obtain the intermediate DW with two equivalent minimum-energy orientations $\bigstar,\bigstar'$ differing by approximately~$90^\circ$.

Figure~\ref{fig:sims} shows how spin waves, generated on the left-hand side of the strip, pass through a DW.
We solve the \LLG{} (LLG) equation, for relaxed initial states, on a square grid ($0.33l\times0.33l$ cells) using a self-developed C++ code \cite{Buijnsters2014} with implicit-midpoint time integration \cite{dAquino2005}.
Spin waves are generated by a space-local, time-periodic in-plane applied field ($\omega=1.70\,|\gamma| K / M_\text{S}$), switched on at $t=0$.
Each waveguide strip ($267l \times 10l$) is simulated in a vacuum-padded periodic box ($333l \times 27l$).
We calculate the difference $\Delta\varphi=\varphi'-\varphi$ in phase shift  between the two chiralities at the right-hand side of the interferometer, comparing intermediate DMI to the case without DMI. A phase difference of up to $180^\circ$ (destructive interference) is obtained for intermediate DMI.

Fixing $4 \pi M_\text{S} = 3.8\text{ kG}$ and $A = 10^{-6}\text{ erg/cm}$ \cite{Emori2013}, the other parameters of Fig.~\ref{fig:sims} become $f = 9.9\text{ GHz}$ (frequency), $W = 127\text{ nm}$ (waveguide width $\approx$ wavelength), $D = 0.047\text{ erg/cm}^2$, and $L = 38\text{ nm}$, assuming the free-electron gyromagnetic ratio.
Spin waves of such frequencies and wavelengths can be experimentally generated, observed, and visualized~\cite{Demidov2008,Ciubotaru2012,Grafe2015}.
Since the DMI is an interfacial effect, $D$ is inversely proportional to film thickness $L$ \cite{Cho2015}. Extrapolation of the values in Ref.~\cite{Emori2013} ($D= 0.5\text{ erg/cm}^2$, $L = 3\text{ nm}$) suggests that $D\sim0.04\text{ erg/cm}^2$ is realistic for $L \sim 38\text{ nm}$.

The phase difference $\Delta\varphi$ between spin waves traveling along the two paths ($\bigstar$ and $\bigstar'$) has a geometric \cite{Dugaev2005} origin. It is convenient to define $\varphi=\varphi_\text{geom}+\varphi_\text{rel}$.
A spin wave causes magnetization to precess around its local equilibrium direction $\mathbf{m}(x)$ \footnote{
In our (semi-)analytical calculations, we assume an infinite waveguide  in the $y$ direction. In the $z$ direction, we consider a finite $L$ in the calculation of the dipolar interactions but neglect the $z$ dependence of the magnetization inside the film, making the magnetization a function of $x$ only.
The two-dimensional simulations of Fig.~\ref{fig:sims} show that the essence of our analytical results carries over to waveguides of finite width $W$.}.
In the limit of exchange spin waves ($k_x\rightarrow\infty$), the dynamics induced by the wave is given by the real part of
\begin{equation}\label{eq:swdef}
\mathbf{m}(x) + \epsilon e^{i[\omega t + k_x x + k_y y + \varphi_\text{rel}(x)]} [ \hat{\mathbf{a}}(x) - i \hat{\mathbf{b}}(x) ]
\text{,}
\end{equation}
where $\epsilon>0$ is the infinitesimal amplitude of the spin wave (linear regime).
The orthonormal basis vectors $\hat{\mathbf{a}}(x), \hat{\mathbf{b}}(x)$ must be perpendicular to $\mathbf{m}(x)$ for all $x$, so that their orientation continually changes across the DW.
A natural choice is to define $\hat{\mathbf{a}}(x), \hat{\mathbf{b}}(x)$ according to parallel transport, $\frac{\ddnoskip{}\hat{\mathbf{a}}}{\ddnoskip{}x} = - (\hat{\mathbf{a}} \cdot \frac{\ddnoskip{}\mathbf{m}}{\ddnoskip{}x}) \mathbf{m}$, by which the basis vectors, at any given point $x$, match their orientation in an infinitesimal neighborhood of $x$ as closely as possible.

The function $\varphi_\text{rel}(x)$ in Eq.~\ref{eq:swdef} determines the phase of the spin wave relative to the basis $\hat{\mathbf{a}}, \hat{\mathbf{b}}$.
However, the orientation of the basis $\hat{\mathbf{a}}, \hat{\mathbf{b}}$ after parallel transportation across the DW strongly depends on the DW configuration ($\bigcirc$, $\bigcirc'$, $\bigstar$, $\bigstar'$, or $\square$), as shown in Fig.~\ref{fig:partrans}(c)--(e). This reorientation of $\hat{\mathbf{a}}, \hat{\mathbf{b}}$ implies an additional phase shift $\varphi_\text{geom}$, which is purely geometric in nature.

\begin{figure}
  \centering
  \includegraphics[scale=1.0]{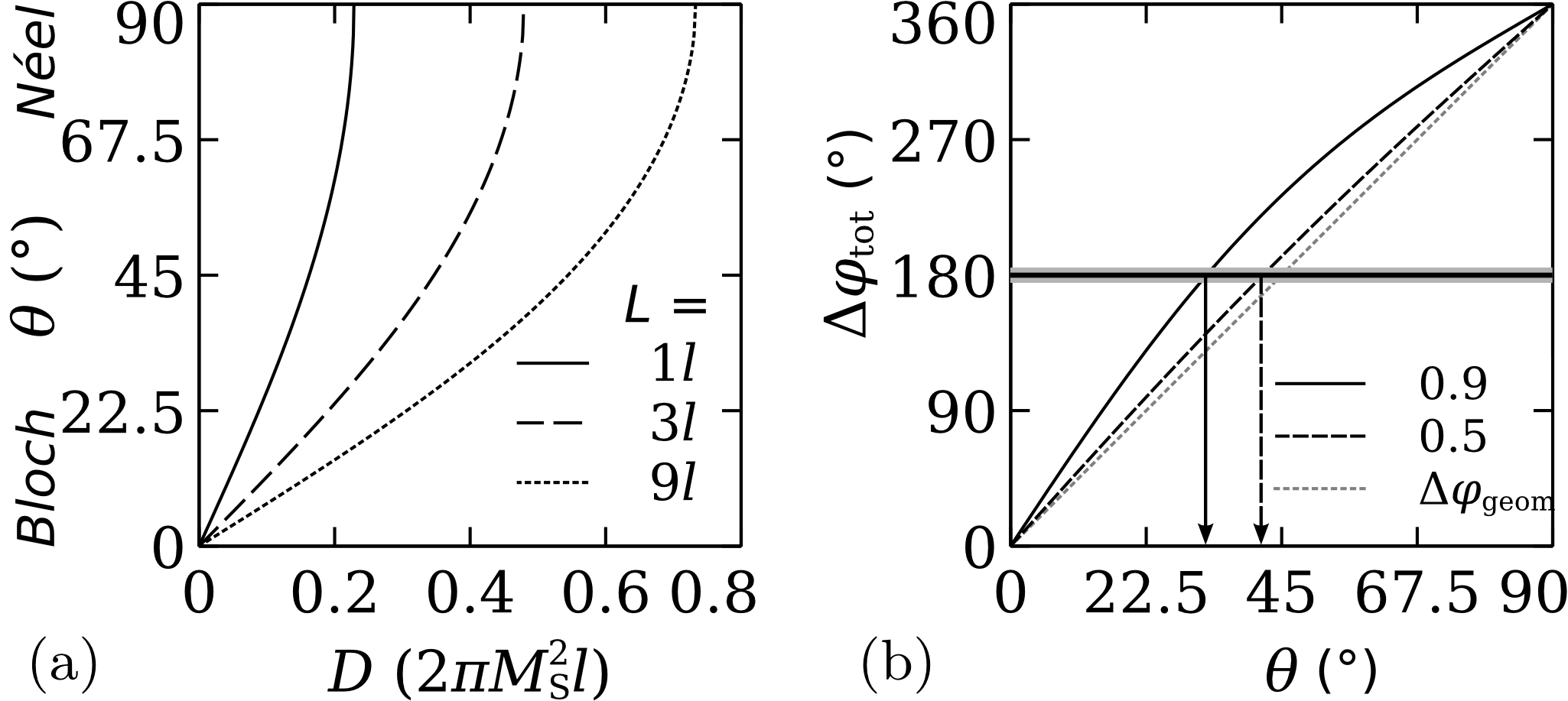}
  \caption{\label{fig:plotsmain}
  (a) DW angle~$\vartheta$ as a function of the ratio of DMI strength $D$ and dipolar interaction, for three  film thicknesses $L$, taking $\sqrt{2\pi M_\text{S}^2 / K} = 0.8$ \cite{capnote1}.
  The angle $\vartheta$ determines whether intermediate DWs ($\bigstar',\bigstar$) are closer to a Bloch or a \Neel{} configuration.
  More ``\Neel{}-like'' DWs (larger $\vartheta$) are obtained for stronger $D$. Conversely, the dipolar interaction penalizes the \Neel{} configuration (this effect is weaker in thinner films).
  (b)~Phase difference $\Delta \varphi$ between spin waves passing through DWs of opposite chiralities ($\bigstar'$ \emph{vs.}\ $\bigstar$), as a function of~$\vartheta$, in the $k_x\rightarrow\infty$ limit, for two values of $\sqrt{2\pi M_\text{S}^2/K}$, taking $L = 3.0 l$.
  The dominant contribution $\Delta\varphi_\text{geom}$ is separated out.
The remaining contribution $\Delta\varphi_\text{rel}$ lowers the value of $\vartheta$ needed for destructive interference ($\Delta\varphi=180^\circ$).
 For $\sqrt{2\pi M_\text{S}^2/K}=0.9$,  $\Delta\varphi_\text{rel}$ is  larger than for $0.5$ because a relatively strong DMI $D /( A w_0^{-1})$ is then needed to obtain a given  $\vartheta$.}
\end{figure}

It is apparent from Fig.~\ref{fig:partrans}(c)--(e) that the geometric contribution is approximately given by
\begin{equation}\label{eq:phigeomtrivial}
\Delta\varphi_\text{geom} \approx 4 \vartheta\text{,}
\end{equation}
where $\vartheta$ is the in-plane angle of the magnetization at the DW center, as shown in Fig.~\ref{fig:partrans}(d).
For example, we have a geometric phase difference $\Delta\varphi_\text{geom} \approx 180^\circ$ for intermediate DWs with $\vartheta = 45^\circ$.
The value of $\vartheta$ is determined by the competition between the DMI and the dipolar interaction, as shown in Fig.~\ref{fig:plotsmain}(a).
While in principle $\Delta\varphi_\text{geom}$ depends on the exact shape of the equilibrium profile $\mathbf{m}(x)$, we find that the deviation from Eq.~\eqref{eq:phigeomtrivial} is at most a few degrees \cite{supmat}.

We derive \cite{supmat} the relative contribution $\Delta\varphi_\text{rel}$ for $k_x\rightarrow\infty$, up to a correction of order $|k_x|^{-1}$, as
\begin{equation}\label{eq:phasedmi}
\Delta\varphi_\text{rel} = \frac{D}{2A} \int_{-\infty}^\infty m'_y(x) \dd{}x -\frac{D}{2A} \int_{-\infty}^\infty m_y(x) \dd{}x\text{,}
\end{equation}
where $m'_y$ and $m_y$ are the magnetization profiles $\bigstar'$,$\bigstar$ calculated numerically, taking into account DMI and dipolar interactions. 
Notice that the exchange interaction does not contribute directly to Eq.~\eqref{eq:phasedmi} because the basis $\hat{\mathbf{a}}(x),\hat{\mathbf{b}}(x)$ (parallel transport) absorbs such a contribution into $\Delta\varphi_\text{geom}$.

Equation~\eqref{eq:phasedmi} gives, approximately,
\begin{equation}\label{eq:phirelapprox}
\Delta \varphi_\text{rel} = \varphi_\text{rel}' - \varphi_\text{rel} \approx \frac{D}{A} w_0 \cos{\vartheta}\text{,}
\end{equation}
where $w_0$ is a characteristic DW width ($w_0\approx \pi l$ for $2\pi M_\text{S}^2 \ll K$).
Notice that $\Delta \varphi_\text{rel}$ vanishes for $D=0$ ($\bigcirc$, $\bigcirc'$) and for large $D$ ($\square$), where $\vartheta = \pi / 2$ (\Neel{} wall).
As shown in Fig.~\ref{fig:plotsmain}(b),
the contribution of $\Delta\varphi_\text{rel}$ enhances the effect of $\Delta\varphi_\text{geom}$ and merely shifts the critical internal angle $\vartheta$ for perfect destructive interference ($\Delta \varphi = 180^\circ$) to a somewhat lower value (more \Bloch{}-like DW).
Therefore the concept of the interferometer spin-wave switch is robust: we
can always find a DW angle $0^\circ < \vartheta < 90^\circ$ such that the phase difference $\Delta \varphi$ between opposite chiralities is exactly $180^\circ$. The desired value $\vartheta$ could then be realized by fine-tuning the DMI strength~$D$ [Fig.~\ref{fig:plotsmain}(a)].

While Eq.~\eqref{eq:phasedmi} is derived in the short-wavelength limit, we have numerically solved the spin-wave normal-mode problem \cite{Buijnsters2014} for incoming waves of arbitrary wavenumber~$k_x$.
The phase shifts $\varphi',\varphi$ depend significantly on $k_x$, as in the case without DMI ($\varphi = 2\arctan{(k_x l)^{-1}}$ \cite{Macke2010}), but the difference $\Delta \varphi = \varphi' - \varphi$ between the two chiralities, which is the relevant quantity in our interferometer, is weakly wavelength dependent for wavelengths comparable to (or shorter than) the DW width.
The weak dependence of $\Delta \varphi$ on $k_x$ can, under certain approximations, also be derived analytically \cite{supmat}.
This observation justifies our approach $k_x \rightarrow \infty$.


In summary, we have shown that the interfacial \DM{} interaction in ultrathin magnetic films provides a new way of manipulation of spin waves. With this interaction, spin waves experience a different phase shift when passing through DWs of different chiralities, leading to either constructive or destructive interference in a two-branch interferometer. One can open or close the transmission of spin waves through the device by changing the DW chirality in one of the two branches. This opens the possibility of developing a memory element or transistor based on the manipulation of magnonic currents without charge transport.


\begin{acknowledgments}
The authors thank K. S. Novoselov for useful discussions.
A. Qaiumzadeh provided some additional literature.
This work is part of the research programme of the Foundation for Fundamental Research on Matter (FOM), which is part of the Netherlands Organisation for Scientific Research (NWO).
Y.F. acknowledges support from the Spanish MECD Grant No. FIS2011-23713.
\end{acknowledgments}

\renewcommand{\theequation}{S\arabic{equation}}
\renewcommand{\thefigure}{S\arabic{figure}}

\appendixheading
\section{\label{sec:dmisyms}Symmetries and the \DM{} interaction}

This section provides some background on the distinct forms of the \DM{} interaction (DMI) referred to in the main text.

\subsection{\label{sec:DMIgen}Bulk versus interfacial DMI}

For the DMI to be noticeable at the continuum level, it is necessary that the central inversion symmetry of the system is broken. This may be the case because the crystal structure itself is noncentrosymmetric (chiral or polar) or because the geometry of the system breaks the inversion symmetry, for instance near a surface or interface.
A variety of continuum models for the DMI exist, which all arise from the same microscopic definition. The difference lies in the orientation of the \DM{} vectors $\mathbf{D}_{ij}$ that define the interaction between atoms $i,j$.

We consider two high-symmetry special cases of the DMI.
The continuum form $E_\text{DMI} = D \int \mathbf{m} \cdot (\nabla\times\mathbf{m}) \dd{}V $ describes a DMI that is \emph{chiral} (it changes sign under any reflection) but which is otherwise completely isotropic.
An alternative form $E_\text{DMI} = -2D \int \mathbf{m} \cdot \nabla(\hat{\mathbf{n}}\cdot\mathbf{m}) \dd{}A $ pertains to a system that is \emph{polar}: the inversion symmetry is broken by a director $\hat{\mathbf{n}}$, which changes sign under central reflection; however, the system is achiral in that it remains invariant under any reflections that leave $\hat{\mathbf{n}}$ intact.
The system is isotropic only in the plane normal to $\hat{\mathbf{n}}$.

The interface of a magnetic layer with another material defines a polar axis $\hat{\mathbf{n}}$ (the interface normal), which gives rise to a DMI of the polar form (if we neglect any in-plane anisotropies). For this reason, the polar form is usually referred to as the interfacial DMI.
The chiral but isotropic form of the DMI is usually referred to as the bulk DMI, since it is the simplest expression that models the DMI in a chiral crystal in bulk (again neglecting any anisotropies). A crystal structure is chiral if it cannot be superimposed onto its mirror image in any way.
We remark, for completeness, that a bulk crystal may also show an `interfacial' DMI effect if it has a polar crystal structure \cite{Kesz2015}.

The distinct effects of the bulk DMI and the interfacial DMI are well known from the theory of magnetic skyrmions, where it is found that the `bulk' model predicts Bloch skyrmions while the `interfacial' model predicts \Neel{} skyrmions \cite{Tomasello2015}.
Similarly, it is found that a strong interfacial DMI tends to push a domain wall into a \Neel{} configuration with a well-defined chirality, while a strong bulk DMI stabilizes one particular chirality of the Bloch domain wall.

In Sec.~\ref{sec:microDMI}, we link the cases of bulk DMI and interfacial DMI to atomistic toy models.
In Sec.~\ref{sec:macroDMI}, we explain that the preference of each type of DMI for one or the other type of domain wall (Bloch \emph{vs.}\ \Neel{}) is a direct consequence of its symmetries.

\subsection{\label{sec:microDMI}Atomistic and continuum models}

\begin{figure}
  \includegraphics[scale=1.0]{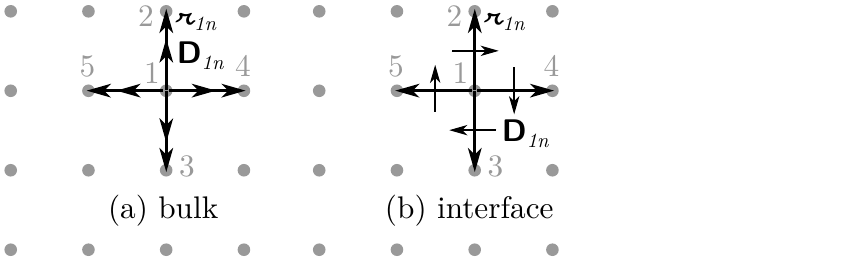}
  \caption{\label{fig:dmvec}
  Directions of the \DM{} vectors $\mathbf{D}_{ij}$ and the lattice displacement vectors $\mathbf{r}_{ij}$ between atoms $i,j$. (a) For a bulk DMI, the $\mathbf{D}_{ij}$ roughly point into the direction of $\mathbf{r}_{ij}$. The $\mathbf{D}_{ij}$ ``diverge away from'' (or ``converge to'') a given atom $i$. (b) For an interfacial DMI, the $\mathbf{D}_{ij}$ ``curl around'' a given atom $i$, with a sense determined by the normal $\hat{\mathbf{n}}$.
  }
\end{figure}

This section shows how the continuum energy functionals for bulk~\eqref{eq:contBulk} and interfacial~\eqref{eq:contInter} DMI can be derived from the atomistic definition~\equaref{eq:DMImicro}, given appropriate choices for the \DM{} vectors $\mathbf{D}_{ij}$.

At the level of individual magnetic moments, the DMI is by definition any interaction that can be written as
\begin{equation}\label{eq:DMImicro}
 E_\text{DMI} = \sum_{ij} \mathbf{D}_{ij} \cdot ( \mathbf{m}_i \times \mathbf{m}_j )
 \text{,}
\end{equation}
where $\mathbf{m}_i$ is the microscopic magnetic moment on site $i$ and $\mathbf{D}_{ij}$ is the \DM{} interaction vector between sites $i$ and $j$, which without loss of generality satisfies $\mathbf{D}_{ij} = -\mathbf{D}_{ji}$.
For simplicity, let us assume that the magnetic atoms of the crystal are arranged as a Bravais lattice, where we use the symbol $\brcurs_{ij} = \mathbf{r}_j - \mathbf{r}_i$ to represent any given (near-neighbor) lattice vector. (Notice that, for obvious reasons, the overall crystal structure of a noncentrosymmetric crystal must be more complicated than just a single Bravais lattice.)

By translation invariance in the bulk, the magnitude and direction of $\mathbf{D}_{ij}$ must be a function of $\brcurs_{ij}$ only. For simplicity, we shall assume that only nearest-neighbor interactions are important and that $\brcurs_{ij}$ represents a nearest-neighbor lattice vector. If the material is isotropic, the only reasonable choice for the \DM{} vector is $\mathbf{D}_{ij} = D \mathbf{\brcurs}_{ij}$, where $D$ is a (positive or negative) interaction strength. The \DM{} vectors $\mathbf{D}_{ij}$ seem to ``diverge'' from any given site $i$, as shown schematically in Fig.~\ref{fig:dmvec}(a).

Near a surface or interface, by contrast, the interface normal $\hat{\mathbf{n}}$ introduces a preferential direction, which we could give a definite sense be defining it to point from material $A$ into material $B$. By symmetry arguments \cite{Crepieux1998}, it is found that the $\mathbf{D}_{ij}$ vectors ``curl around'' a given site $i$ with a well-defined sense induced by the direction of $\hat{\mathbf{n}}$, as shown schematically in Fig.~\ref{fig:dmvec}(b). Considering only the highest layer of magnetic atoms below the interface and only their nearest-neighbor interactions, the only natural choice in the absence of in-plane anisotropy is $\mathbf{D}_{ij} = D \hat{\mathbf{n}} \times \mathbf{\brcurs}_{ij}$, since $\mathbf{D}_{ij}$ must be a vector-valued function linear both in $\mathbf{\brcurs}_{ij}$ (because of the rule $\mathbf{D}_{ij} = -\mathbf{D}_{ji}$) and in $\hat{\mathbf{n}}$ (because the polar axis $\hat{\mathbf{n}}$ is the only element that breaks inversion symmetry).

Passing to a continuum theory, Eq.~\eqref{eq:DMImicro} becomes
\begin{equation}\label{eq:DMImacrogen}
E_\text{DMI} = \int \sum_{\brcurs} \mathbf{D}(\rcurs) \cdot [\mathbf{m} \times (\brcurs \cdot \nabla) \mathbf{m}] \, \Omega^{-1} d^3 r
\text{,}
\end{equation}
where $\Omega$ represents a unit-cell volume and $\brcurs$ sums over all relevant near-neighbor lattice vectors.
In the case of isotropic bulk DMI with only nearest-neighbor interactions, we substitute $ \mathbf{D}(\brcurs) = D\brcurs$. Equation~\eqref{eq:DMImacrogen} becomes, in tensor notation,
\begin{equation}
E_\text{DMI} = \int D \varrho_{ad} \epsilon_{abc} m_b  \partial_d m_c  \, \Omega^{-1} d^3 r
\text{,}
\end{equation}
where $\varrho_{ab} = \sum_{\brcurs} \rcurs_a \rcurs_b $.
Assuming $\varrho_{ab} = \alpha\delta_{ab}$ for some scalar $\alpha>0$ (isotropy), we get
\begin{equation}\label{eq:contBulk}
E_\text{DMI} = - \alpha D \int m_b \epsilon_{bac} \partial_a m_c  = -\alpha D \int \mathbf{m} \cdot (\nabla \times \mathbf{m}) \text{.}
\end{equation}

For the interfacial DMI, on the other hand, we have $\mathbf{D}_{ij} = D \hat{\mathbf{n}} \times \brcurs_{ij}$, where $\hat{\mathbf{n}}$ is the interface normal. For simplicity, we take $\hat{\mathbf{n}} = \hat{\mathbf{z}}$.
We get
\begin{subequations}
\begin{align}
E_\text{DMI}& =
\sum_{\brcurs} \int (D\epsilon_{abc} z_b \rcurs_c) \epsilon_{ade} m_d (\rcurs_f \partial_f m_e)
\, \Omega^{-1} d^2 r \\
& = \int D \varrho_{cf} \epsilon_{abc} \epsilon_{ade} z_b m_d  \partial_f m_e \, \Omega^{-1} d^2 r \\
& = \int D \varrho_{cf} (\delta_{bd}\delta_{ce} - \delta_{be}\delta_{cd}) z_b m_d  \partial_f m_e \, \Omega^{-1} d^2 r
\text{.}
\end{align}
\end{subequations}
Notice that we now integrate over the interface plane only and assume that $\mathbf{m}$ does not depend on the perpendicular coordinate $z$ (here $\Omega$ represents a unit area). Assuming $\varrho_{ab} = \alpha\delta_{ab}$, we get
\begin{subequations}
\begin{align}
E_\text{DMI}& = \alpha D
\int (
  z_b m_b  \partial_c m_c
- 
  z_b m_c  \partial_c m_b 
)\,\Omega^{-1} d^2 r \\
& = \alpha D
\int [
  (\hat{\mathbf{z}}\cdot \mathbf{m})  (\nabla \cdot \mathbf{m})
- 
  \mathbf{m} \cdot \nabla (\hat{\mathbf{z}} \cdot \mathbf{m})
]\,\Omega^{-1} d^2 r
\text{.}
\end{align}
\end{subequations}
Using the divergence theorem, we get
\begin{equation}\label{eq:contInter}
E_\text{DMI} = 
- \frac{2\alpha D}{\Omega} \int_U \mathbf{m} \cdot  (\nabla m_z)  \dd^2 r
+ \frac{\alpha D}{\Omega} \oint_{\partial U} (m_z\mathbf{m}) \cdot \ddnoskip{}\mathbf{z}_{\partial U}   \text{,}
\end{equation}
where $m_z(r) = \hat{\mathbf{z}} \cdot \mathbf{m}(r)$.
The boundary term is irrelevant if we integrate over all space ($U=\mathbb{R}^2$).

\subsection{\label{sec:macroDMI}Effect on domain-wall energy}

\begin{figure}
  \centering
  \includegraphics[scale=1.0]{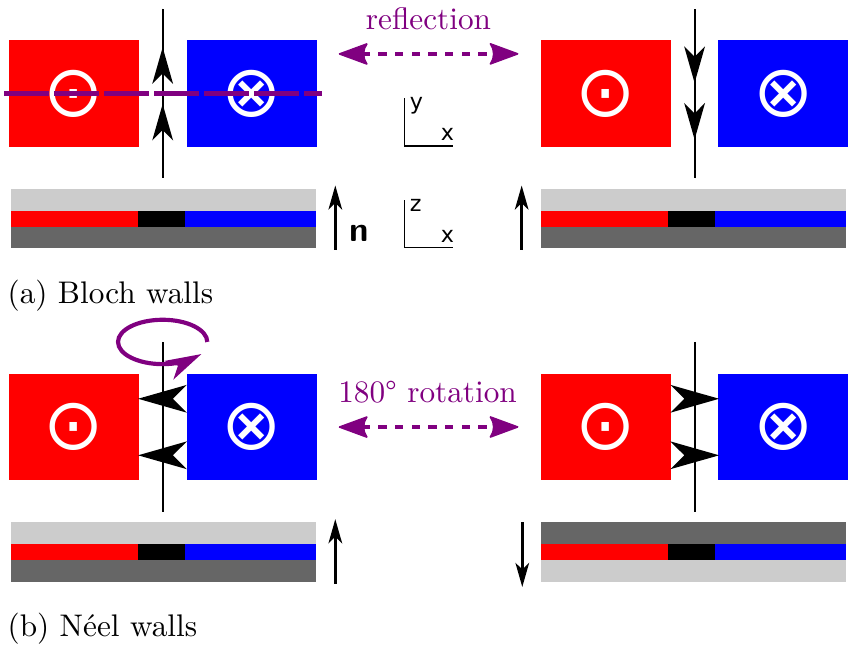}
  \caption{\label{fig:dwsym}(color online).
  Symmetries of the two types of domain wall.
  (a) The two distinct states of the Bloch wall are related by a reflection that leaves the surface normal $\hat{\mathbf{n}}$ invariant.
  The bulk DMI selects one Bloch state as the most favorable and the other Bloch state as the least favorable orientation. The interfacial DMI does not discriminate between the two.
  (b) The two distinct states of the \Neel{} wall are related by $180^\circ$ rotation
  that reverses the surface normal $\hat{\mathbf{n}}$.
  The interfacial DMI selects one \Neel{} state as the most favorable and the other \Neel{} state as the least favorable orientation. The bulk DMI does not discriminate between the two.
  }
\end{figure}

While the bulk DMI is chiral but isotropic, the interfacial DMI is polar but achiral (and isotropic under rotations around the $\hat{\mathbf{z}}$ axis).
The two forms considered here are the only possibilities that satisfy these respective constraints, regardless of the details of the microscopic model.
The same symmetry considerations determine how the two forms of the DMI respond to Bloch or \Neel{} domain walls in a thin film with perpendicular anisotropy.
In Fig.~\ref{fig:dwsym}, we compare the two chirality states that exist of either type of wall.
The bulk DMI selects one of the two chiralities of the Bloch wall as its most favorable orientation (energy minimum), and the other Bloch chirality as the least favorable orientation (energy maximum). On the other hand, the interfacial DMI selects one of the two chiralities of the \Neel{} wall as its most favorable orientation, and the other \Neel{} chirality as the least favorable orientation.

We can interpret this behavior as follows.
The two chirality states of the Bloch wall are related by a reflection in the $xz$ plane. Since such a reflection leaves the $z$-director unchanged, it follows that the interfacial DMI cannot discriminate between the two chiralities of the Bloch wall. The bulk DMI can, because the bulk DMI is not invariant under any spatial reflections.
The two chiralities of the N\'eel wall, on the other hand, are related by a $180^\circ$ rotation around the $y$ axis. Such a rotation is a symmetry of the bulk DMI (isotropy) but not of the interfacial DMI (reversal of $\hat{\mathbf{n}}$). As a consequence, only the interfacial DMI discriminates between the two chiralities of the \Neel{} wall.

The orientation of a Bloch domain wall will not be affected by a small bulk DMI, as both chiralities represent extrema of the bulk DMI. A small interfacial DMI, by contrast, causes a change in orientation of the Bloch domain wall, reorienting it slightly towards the favorable \Neel{} state.

\section{Dipolar interaction}

This section specifies the energy functionals used to describe the dipolar (magnetostatic) interaction in our micromagnetic simulations and numerical calculations.

Throughout this work, we assume that the magnetization is homogeneous in $z$ inside the ferromagnetic film (uniform-mode approximation). In other words, we assume that the magnetization may be written as $\mathbf{M}(x,y,z) = M_\text{S} \mathbf{m}(x,y,z) = M_\text{S}  \Pi(z/L) \mathbf{m}(x,y)$, where $L$ is the film thickness and where $\Pi(\xi) = 1$ for $|\xi| < \frac{1}{2}$ and $\Pi(\xi) = 0$ otherwise.
This simplification can be justified in the regime that $L \lessapprox l$, where $l$ is the exchange length.

It can be derived (see, for example, Ref.~\cite{Buijnsters2015B}) that the total dipolar energy (defined per unit thickness) is given, in reciprocal space, by
\begin{equation}\label{eq:gendipint}
E_\text{dip} = 2 \pi M_\text{S}^2 \int \tilde{m}^*_a(\mathbf{k}) \tilde{g}_{ab}(\mathbf{k}) \tilde{m}_b(\mathbf{k}) \frac{\ddnoskip^2k}{(2\pi)^{2}}
\end{equation}
with
\begin{subequations}\label{eq:greci}
\begin{align}
\tilde{g}_{uv}(k_x,k_y)& = (1-N_k) \frac{k_u k_v}{k^2} \text{,}\\
\tilde{g}_{uz}(k_x,k_y)& = 0\text{,}\\
\tilde{g}_{zz}(k_x,k_y)& = N_k \text{,}
\end{align}
\end{subequations}
where $k= \sqrt{k_x^2 + k_y^2}$ and where the indices $u,v$ represent the in-plane coordinates $x,y$. 
The function $m_a(\mathbf{r})$ describes the $a$ component of the normalized magnetization field. 
Its Fourier transform $\tilde{m}_a(\mathbf{k})$ is defined as
\begin{equation}
\tilde{m}_a(\mathbf{k}) = \iint m_a(\mathbf{r}) e^{-i\mathbf{k}\cdot\mathbf{r}} \dd^2\mathbf{r}
\text{.}
\end{equation}
The demagnetizting factor $N_k$ is given by
\begin{equation}
N_k = \frac{1-e^{-kL}}{kL}\text{.}
\end{equation}
We define $N_{k=0} = 1$ by continuity. Notice that $N_{k\rightarrow\infty} = 0$.
We have omitted from Eq.~\eqref{eq:gendipint} some contributions of the form  $\iint \lVert \tilde{\mathbf{m}}(\mathbf{k}) \rVert^2 \ddnoskip^2k = (2\pi)^{2} \iint \lVert \mathbf{m}(\mathbf{r}) \rVert^2 \dd^2\mathbf{r}$, which are constant. 

For a system with periodic boundary conditions that is defined on a rectangular grid, 
we evaluate the dipolar interaction in $\mathcal{O}(n \log n)$ time using a fast Fourier transform (FFT).
Finite magnetic elements, such as a waveguides with a finite width, might also, somewhat approximately, be simulated in this way, provided that sufficient zero padding is inserted around the object to minimize the influence of the other periodic copies.

For a one-dimensional magnetization profile $\mathbf{m}(x)$, assumed to extend infinitely and uniformly in the $y$ direction ($k_y=0$), we may write the dipolar energy in real space as
\begin{equation}\label{eq:dip1D}
E_\text{dip} = 2 \pi M_\text{S}^2 \iint m_a(x') g_{ab}(x'-x) m_b(x) \dd{}x \dd{}x'
\end{equation}
where
\begin{subequations}
\begin{align}
\label{eq:dipxxpart}g_{xx}(x)& = \delta(x)  -  \frac{1}{2\pi L} \log\Bigl( 1 + \frac{L^2}{x^2} \Bigr) \text{,} \\
g_{zz}(x)& = \frac{1}{2\pi L} \log\Bigl( 1 + \frac{L^2}{x^2} \Bigr) \text{,} \\
g_{xy}(x)& = g_{yy}(x) = g_{xz}(x) = g_{yz}(x) = 0 \text{.}
\end{align}
\end{subequations}
These expressions can be useful when evaluating the dipolar interaction for a system that is not periodic in $x$, such as a domain wall, where we have $\mathbf{m}(-\infty) = \hat{\mathbf{z}}$ while $\mathbf{m}(\infty) = -\hat{\mathbf{z}}$.
We remark that an efficient numerical implementation of Eq.~\eqref{eq:dip1D} might still evaluate the convolution through an FFT, but special care must be taken to correctly take into account the asymptotic behavior of $\mathbf{m}(x)$ for $x\rightarrow-\infty$ and $x\rightarrow\infty$.

\section{\label{sec:phWKB}Phase shift in the WKB approximation ($k\rightarrow\infty$)}

\begin{figure}
  \includegraphics[scale=1.0]{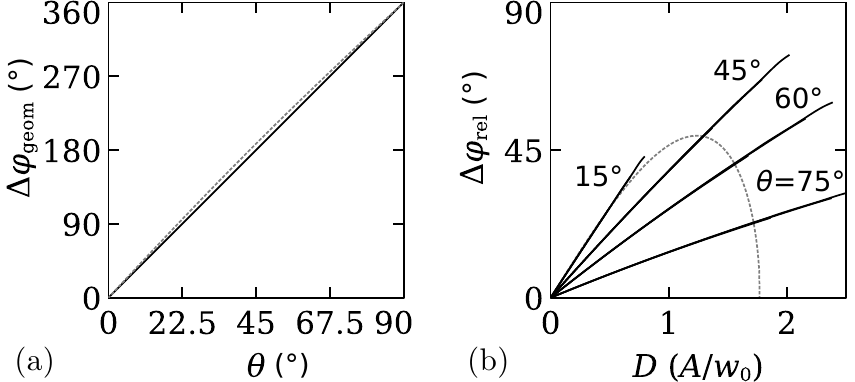}
  \caption{\label{fig:plotssup}
  Phase-shift contributions $\Delta \varphi_\text{geom}$ and $\Delta \varphi_\text{rel}$ (for $k_x\rightarrow\infty$) obtained from the minimum-energy domain-wall profiles, which are numerically calculated as a function of the parameters $A$, $K$, $M_\text{S}$, $L$, and $D$ taking the dipolar interaction~\eqref{eq:dip1D} into account. The variables $\vartheta$ and $w_0$ are also evaluated.
  (a)~The geometric part $\Delta \varphi_\text{geom}$ follows Eq.~(3) of the main text (solid line), with a deviation of at most a few degrees even in extreme cases (dashed line, $\sqrt{2\pi M_\text{S}^2 / K} = 0.98$).
  (b)~The relative part $\Delta \varphi_\text{rel}$ depends on the DMI strength $D$ (as compared to exchange) and on the domain-wall angle $\vartheta$; it follows approximately Eq.~(5) of the main text. For given $Dw_0 / A$ and $\vartheta$, the dependence on the third dimensionless parameter is negligible.
  The dashed line indicates the relation between $D$ and $\vartheta$ if we fix $\sqrt{2\pi M_\text{S}^2/K} = 0.90$ and $L = 3.0l$.
  }
\end{figure}

In this section, we provide a derivation of our expression for the relative part $\varphi_\text{rel}$ of the phase shift in the $k\rightarrow\infty$ limit, presented in the main text as Eq.~(4). In \figref{fig:plotssup}, we evaluate the geometric phase shift $\varphi_\text{geom}$ and the expression~(4) for $\varphi_\text{rel}$ for numerically calculated equilibrium domain-wall profiles $\mathbf{m}(x)$, confirming the accuracy of Eqs.~(3) and~(5) of the main text.

An equilibrium magnetization profile $\mathbf{m}(x)$ is found by minimization of total energy $E$ under the constraint that $\lVert\mathbf{m}(x)\rVert=1$ for all $x$.
It satisfies
\begin{equation}\label{eq:equiconf}
\frac{\delta E}{\delta \mathbf{m}(x)} = -h(x) \mathbf{m}(x) \text{,}
\end{equation}
where the expression on the left-hand size denotes a functional derivative. The scalar-valued function $h(x)$ is a Lagrange multiplier.
The equilibrium profile $\mathbf{m}(x)$ suffices to calculate the geometric phase induced by parallel transport of the basis vectors.
However, to determine the relative phase we must solve 
the normal-mode equation, which can be expressed in coordinate-free form [23] as
\begin{multline}\label{eq:normalmode}
\int \frac{\delta^2 E}{\delta \mathbf{m}(x) \delta\mathbf{m}(x')} \cdot \mathbf{u}(x') \dd{}x' + h(x) \mathbf{u}(x) \\
 - \frac{M_\text{S}}{|\gamma|}i\omega[\mathbf{m}(x)\times \mathbf{u}(x)] = \lambda(x) \mathbf{m}(x)\text{,}
\end{multline}
where $\gamma$ is the gyromagnetic ratio, $h(x)$ is fixed by Eq.~\eqref{eq:equiconf}, and where $\lambda(x)$ is a Lagrange multiplier,
under the constraint that $\mathbf{m}(x)\cdot\mathbf{u}(x)=0$ at all $x$.
The frequency $\omega$ is fixed by the wavenumber $k_x$ of the spin wave far away from the domain wall.

Equation~\eqref{eq:normalmode}, when written out explicitly in terms of $\mathbf{u}(x) = a(x)\hat{\mathbf{a}}(x) + b(x)\hat{\mathbf{b}}(x)$, becomes a complicated integro-differential equation. (Nonlocality arises from the dipolar interaction.)
Here we find an approximate solution using the \WKB{} (WKB) approximation, which is exact in the short-wavelength limit.
We assume that the solution $\sim e^{i \int^x k'_x(x') \ddsmall{}x'}$ locally resembles a plane wave and replace any differential operator $\partial_x$ with $i k'_x$ (analogously for convolution operators) in order to solve for $k'_x$ independently for each $x$.
For the interactions specified in the main text, including interfacial DMI, we get
\begin{equation}\label{eq:WKBmatrix}
\left(
\begin{array}{c c}
A (k'_x)^2 & \frac{M_\text{S}\omega}{2|\gamma|}i + Dm_y ik'_x \\
-\frac{M_\text{S}\omega}{2|\gamma|}i - Dm_y ik'_x & A (k'_x)^2
\end{array}
\right)
\left(
\begin{array}{c}
a \\
b
\end{array}
\right)
= 0
\text{,}
\end{equation}
where, anticipating the limit $|k_x|\rightarrow\infty$, we have written only terms that are of at least first order in $k'_x$ or $k_x$.
The exchange interaction acts as a scalar $A (k_x')^2$ because we define the basis $\hat{\mathbf{a}}(x),\hat{\mathbf{b}}(x)$ according to parallel transport.
We substitute the dispersion relation far away from the domain wall, $\omega \approx (2|\gamma| A/ M_\text{S}) k_x^2$, again up to corrections of constant order.
The characteristic equation of the matrix in Eq.~\eqref{eq:WKBmatrix} has four solutions. We are interested in the solution $k_x'$ closest to $k_x$, which is given by
\begin{equation}\label{eq:deltakx}
k'_x(x) = k_x + \frac{D}{2A} m_y(x) + \mathcal{O}(|k_x|^{-1})\text{.}
\end{equation}
The phase induced by the domain wall on top of the phase factor $e^{i k_x x}$ is now given in the WKB approximation by $\varphi = \int_{-\infty}^\infty [k'_x(x) - k_x] \ddsmall{}x$. We find
\begin{equation}\label{eq:phasedmisup}
\varphi_\text{rel} = \frac{D}{2A} \int_{-\infty}^\infty m_y(x) \dd{}x \text{,}
\end{equation}
up to a correction of order $|k_x|^{-1}$.

Figure~\ref{fig:plotssup} shows a numerical evaluation of the two contributions $\Delta \varphi_\text{geom}$ and $\Delta \varphi_\text{rel}$ to the phase-shift difference $\Delta \varphi = \varphi' - \varphi$ between domain walls of opposite chirality (\emph{e.g.}, $\bigstar'$ and $\bigstar$ in the main text).
To describe the behavior of $\Delta \varphi_\text{rel}$, we introduce the characteristic domain-wall width $w = \int \sqrt{m_x^2 + m_y^2} \dd{} x$, which we evaluate from the equilibrium profile $\mathbf{m}(x)$. The variable $w_0$ is then the value of $w$ for an equivalent domain wall with $D=0$ and the other parameters ($A$, $K$, $M_\text{S}$, and $L$) the same.

For completeness, we derive that the equivalent of \equaref{eq:phasedmisup} for a bulk DMI is $\varphi_\text{rel} = -[D/(2A)] \int_{-\infty}^\infty m_x(x) \ddsmall{}x$. Notice that the latter expression vanishes for any Bloch domain wall.

\section{\label{sec:phloc}Phase shift in a localized model}

In this section, we derive analytically, under certain approximations, the effect of an interfacial DMI on the equilibrium profile and phase shift of a Bloch domain wall.
In particular, we present a closed-form expression~\eqref{eq:phasediff} for the phase shift that is valid for arbitrary wavenumber $k_x$.
The expression suggests that the difference in phase shift between the two chiralities is almost independent of the wavelength of the spin wave.

Our analytical treatment complements the WKB approach of \secref{sec:phWKB} in two ways. First, it allows one to obtain a semianalytical expression for the equilibrium domain-wall profile, which in the WKB approach is taken as given (ie, calculated numerically). Second, it provides an expression for the phase shift for arbitrary $k_x$, where the WKB approach considers only the $k_x\rightarrow\infty$ limit.
The two approximations made are that we take into account the main effect of the dipolar interaction as an effective local interaction, and that we treat the effect of the DMI perturbatively (small $D$).
The results presented here are consistent with the WKB expression (\secref{sec:phWKB}) if we substitute into \eqref{eq:phasedmisup} the equilibrium profile $\mathbf{m}(x)$ calculated for the approximated dipolar interaction [but notice that we need to add the geometric part $\varphi_\text{geom}$ to \equaref{eq:phasedmisup} to obtain the total phase shift].

\subsection{Simplified treatment of dipolar interaction}

In our analytical treatment, we follow \explcite{Shibata2011} in including the dipolar effects (shape anisotropy) into a local anisotropy energy.
We argue that, on scales much smaller than the film thickness $L$, only the $\delta$-function part of Eq.~\eqref{eq:dipxxpart} is important and the effect of the dipolar interaction~\eqref{eq:dip1D} reduces to a local anisotropy $\int K_\perp m_x(x)^2 \dd{}x$ with $K_\perp = 2\pi M_\text{S}^2$. Notice that, even though the dipolar interaction is isotropic, the $x$ coordinate plays a special role because we assume that $\hat{\mathbf{x}}$ is the domain-wall normal (magnetization is a function of $x$ only). The total anisotropy energy is now given by
\begin{equation}\label{eq:ani}
E_\text{ani} = \int \bigl(-K m_z^2 + K_\perp m_x^2\bigr) \dd{}x
\text{,}
\end{equation}
where $K,K_\perp$ are positive constants. The in-plane anisotropy $K_\perp$ models the dipolar interaction in introducing a preference for Bloch domain walls (flux closure).

\begin{figure}
  \includegraphics[scale=1.0]{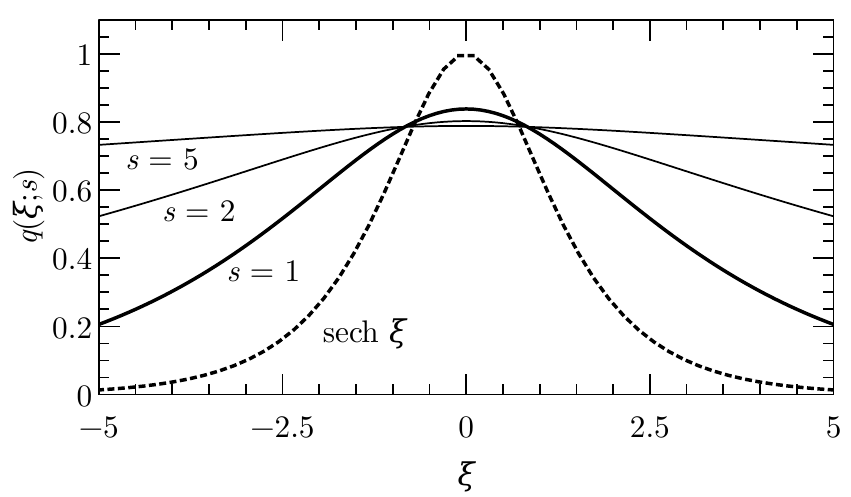}
  \caption{\label{fig:funcg}
    The function $q(\xi;s)$, defined by \equaref{eq:deffuncg}, for several values of the parameter $s$. The special function $\sech \xi$ is shown for comparison.
  }
\end{figure}

This localized approximation is formally valid in the limit that the magnetization profile $\mathbf{m}$ depends only on $x$ and extends infinitely not only in the $y$ but also in the $z$ direction. In practice, this means that we assume normal incidence of spin waves and a wavelength and exchange length much shorter than film thickness $L$.
While these assumptions are unrealistic in most practical cases, they allow us to obtain some   analytical results that are qualitatively correct.

In addition to the easy-axis and in-plane anisotropies, we take into account the interfacial DMI
\begin{equation}\label{eq:intDMI}
 E_\text{DMI} = -2D \int m_x(x)  m_z'(x) \dd{}x
\end{equation}
and the usual exchange term $ E_\text{ex} = A \int \lVert \mathbf{m}'(x) \rVert^2 \dd{}x $, where a prime denotes a derivative with respect to $x$.

\subsection{Equilibrium profile}

In a one-dimensional system, the magnetization profile may be described by two functions $\theta(x),\phi(x)$, defined by 
\begin{equation}
\mathbf{m}(x) 
 = \left(
 \begin{array}{c}
 m_x \\
 m_y \\
 m_z
 \end{array}
 \right)
 = \left(
 \begin{array}{c}
 \sin\theta\cos\phi \\
 \sin\theta\sin\phi \\
 \cos\theta
 \end{array}
\right) \text{.}
\end{equation}
The equilibrium magnetization profile of the domain wall is a solution of
\begin{subequations}\label{eq:varderiv}
\begin{align}
\nonumber
\frac{\delta E}{\delta \theta(x)} &= - 2A\theta'' + A(\phi')^2 \sin2\theta \\\nonumber
 &\phantom{={}}+  ( K + K_\perp \cos^2\phi ) \sin2\theta \\
 &\phantom{={}}- 2D \sin^2\theta\sin\phi \phi' = 0\text{,} \\
\nonumber 
\frac{\delta E}{\delta \phi(x)} &= 2A\sin\theta \phi'' + 4A\cos\theta \theta'\phi' \\\nonumber 
 &\phantom{={}}+ 2K_\perp \sin\theta\sin\phi\cos\phi \\
 &\phantom{={}} -2D \sin\theta\sin\phi \theta'  = 0 \text{,}
\end{align}
\end{subequations}
where the expressions on the left-hand side are functional derivatives of total energy.

For a system with only exchange $A$ and uniaxial anisotropy $K$ ($D=K_\perp=0$), it is well known that the equilibrium profile, assuming $\mathbf{m}(-\infty) = \hat{\mathbf{z}}$ and $\mathbf{m}(\infty) = -\hat{\mathbf{z}}$, is given by
\begin{equation}\label{eq:DWoriginal}
\theta_0(x) = 2\arctan[\exp(x/l)] \text{,}
\end{equation}
where $l = \sqrt{A/K}$ is the exchange length,
while the function $\phi_0(x)$ takes an arbitrary constant value.
If we set $K_\perp>0$, we get the well-known Bloch magnetization profile
\begin{subequations}\label{eq:DWprofileunpert}
\begin{align}
\theta_0(x)& = 2\arctan[\exp(x/l)]\text{,}\\
\phi_0(x)& = \pm \pi / 2 \text{.}
\end{align}
\end{subequations}
The sign ($\pm$) defines the chirality of the domain wall. Both chiralities represent equivalent stable energy minima.
The positive sign corresponds to the configuration $\bigcirc'$ as defined in the main text; the negative sign to $\bigcirc$. 

As a result of the competition with $K_\perp$, the interfacial DMI~\eqref{eq:intDMI} modifies the domain-wall profile, as calculated numerically in \explcite{Heide2006}.
Here, we treat the DMI as a small perturbation ($|D| \ll K_\perp l$).
It can be derived that we get, to first order in $D$, a minimum-energy configuration
\begin{subequations}\label{eq:DWprofilepert}
\begin{align}
\theta_0(x)& = 2\arctan[\exp(x/l)]\text{,}\\
\phi_0(x)& = \pm \pi / 2 \pm \frac{D}{K_\perp l} q\left(\frac{x}{l};s\right) \text{,}&
\end{align}
\end{subequations}
where $s=\sqrt{K/K_\perp}$. Perpendicular magnetization implies $s>1$.
The function $q(\xi;s)$, shown in \figref{fig:funcg}, is uniquely defined as the solution of
\begin{equation}\label{eq:deffuncg}
\left( -s^2 \cosh^2\xi \frac{d}{d\xi} \sech^2\xi \frac{d}{d\xi} + 1 \right)q = \sech\xi
\end{equation}
that is even and vanishes at infinity (particular part).

It is useful to compare \equaref{eq:DWprofilepert} to the equilibrium profile that is obtained if a bulk DMI instead of the interfacial DMI is present.
In that case, we obtain the same equilibrium profile as in \equaref{eq:DWprofileunpert}; in other words, the bulk DMI has no effect on the profile of the Bloch domain wall.
The reason is that, given \emph{any} Bloch profile, where $\phi(x) = \pm \pi / 2$ and $\theta(x)$ is arbitrary, 
the functional derivatives $\delta E_\text{DMI} / \delta \theta(x)$  and $\delta E_\text{DMI} / \delta \phi(x)$ of the bulk DMI energy $E_\text{DMI} = D \int \mathbf{m} \cdot (\nabla \times \mathbf{m}) \ddsmall{}x = -2D \int m_y m'_z\,dx$ with respect to the profile functions $\theta(x),\phi(x)$ vanish.

\subsection{Linearized dynamics}

The dynamics is described by the \LLG{} (LLG) equation without damping
\begin{equation}
-\frac{M_\text{S}}{|\gamma|} \frac{\partial \mathbf{m}}{\partial t} = \mathbf{m} \times \left(-\frac{\delta E}{\delta \mathbf{m}(x,y)} \right)\text{,}
\end{equation}
where $M_\text{S}$ is the saturation magnetization, $\gamma$ is the gyromagnetic ratio, and $E = E_\text{ex} + E_\text{ani} + E_\text{DMI}$ is the total interaction energy.
Let us consider small variations $\delta\theta = \theta(t,x,y) - \theta_0(x)$ and $\delta\phi = \phi(t,x,y) - \phi_0(x)$ around \equaref{eq:DWprofilepert}.
We get a linearized equation of motion
\begin{widetext}
\begin{equation}\label{eq:lineom}
\left( \begin{array}{ccc}
-A \left( \partial_x^2 + \partial_y^2 \right) + K(1-2\sech^2 x/l) & 
-{[}M_\text{S}/(2|\gamma|)]\partial_t   \mp (2D/l) \hat{F} \\
{[}M_\text{S}/(2|\gamma|)]\partial_t    \mp (2D/l) \hat{F}^\dagger & 
- A\left( \partial_x^2 + \partial_y^2 \right) + K(1-2\sech^2 x/l) + K_\perp
\end{array} \right)
\left( \begin{array}{c c}
\delta \theta \\
\sech (x/l)\delta \phi
\end{array} \right)
=
0
\end{equation}
\end{widetext}
where
\begin{multline}
\hat{F} = (l/2)\sech(x/l) h(x/l;s) \partial_x \cosh(x/l)\\
 + q(x/l;s)\tanh(x/l)
\end{multline}
with $h = s^2 q'' - q$
is the perturbation caused by the DM interaction, to first order in $D$.
Notice that $\hat{F}$ is odd in $x$ and breaks reflection symmetry. (The system remains invariant under a simultaneous reflection and change in polarity, which is equivalent to a rotation around $\hat{\mathbf{z}}$.)

\subsection{Transmission phase shift}

We assume that the solutions are periodic in $t$ and $y$.
\Equaref{eq:lineom} becomes a Hamiltonian normal-mode problem \cite{Buijnsters2014}.
Away from the domain wall ($|x| \gg l$), the spin waves take the form 
\begin{equation}\label{eq:scatstates}
\left( \begin{array}{c c}
\delta \theta \\
\sech (x/l)\delta \phi
\end{array} \right)
=
\left( \begin{array}{c c}
1 \\
-i \sqrt{\frac{K + Ak^2}{K + K_\perp + Ak^2}}
\end{array} \right) e^{i(\omega t +  k_x x +  k_y y)}
\end{equation}
with $\omega = (2|\gamma|/M_\text{S}) \sqrt{(K + Ak^2)(K + K_\perp + Ak^2)}$, where $k = \sqrt{k_x^2+k_y^2}$.

We now calculate the reflection ($r$) and transmission ($t$) amplitudes of an incoming spin wave with wavenumber $k_x$ that propagates in the positive-$x$ direction.
The amplitudes $r',t'$ refer to an incoming spin wave wave that propagates in the negative-$x$ direction (wavenumber $-k_x$) and approaches the domain wall from the other side.
\Equaref{eq:scatstates} defines the scattering states.

For $D = 0$, the analytic solutions of \equaref{eq:lineom} are well known \cite{Helman1991}. The propagating-wave solutions take the form
\begin{equation}
\left( \begin{array}{c c}
1 \\
-i \sqrt{\frac{K + Ak^2}{K + K_\perp + Ak^2}}
\end{array} \right) f(x) e^{i(\omega t +  k_x x +  k_y y)}\text{,}
\end{equation}
where \cite{Macke2010}
\begin{equation}
f(x) = - i k_x l + \tanh (x/l)
\text{.}
\end{equation}
It follows immediately that
\begin{equation}
\left( \begin{array}{ccc}
r & t'\\
t & r'
\end{array} \right)
= 
\left( \begin{array}{ccc}
0 & e^{i\varphi_0}\\
e^{i\varphi_0} & 0
\end{array} \right)
\text{,}
\end{equation}
where 
\begin{equation}\label{eq:phi0basic}
\varphi_0 = \arg \frac{(i k_x l - 1)}{(i k_x l + 1)} = 2\arctan \frac{1}{k_x l} \text{.}
\end{equation}
Notice that the domain wall shows total transmission in the localized model.
[However, if the full dipolar interaction is taken into account, we find a nonzero reflection for long-wavelength spin waves (see also, for example, \cite{Macke2010}).]

We now calculate the effect of the interfacial DMI on the transmission phase and amplitude.
As above, we take into account the DMI to first order in the interaction strength $D$.
With some algebraic work, we obtain
\begin{equation}
\left( \begin{array}{ccc}
r & t'\\
t & r'
\end{array} \right)
=
\left( \begin{array}{ccc}
0 & e^{i(\varphi_0 - \varphi_1)}\\
e^{i(\varphi_0 + \varphi_1)} & 0
\end{array} \right)
\text{,}
\end{equation}
where $\varphi_0$ is again given by \equaref{eq:phi0basic}, and where 
\begin{multline}\label{eq:phasediff}
\varphi_1 = \mp \pi \left( \frac{D}{K l} + \frac{D}{K_\perp l} \right)\\
\times \frac{\sqrt{(K + Ak^2)(K + K_\perp + Ak^2)}}{2K + K_\perp + 2Ak^2}\text{.}
\end{multline}
Notice that only $\varphi_1$ depends on chirality ($\pm$).
Since \eqref{eq:lineom} defines the scattering problem relative to the basis vectors $\hat{\theta}, \hat{\phi}$, which have a fixed orientation at $x=-\infty$ and at $x=\infty$, the expression~\eqref{eq:phasediff} includes both the geometric and relative parts of the phase shift.

\begin{figure}[b]
  \includegraphics[scale=1.0]{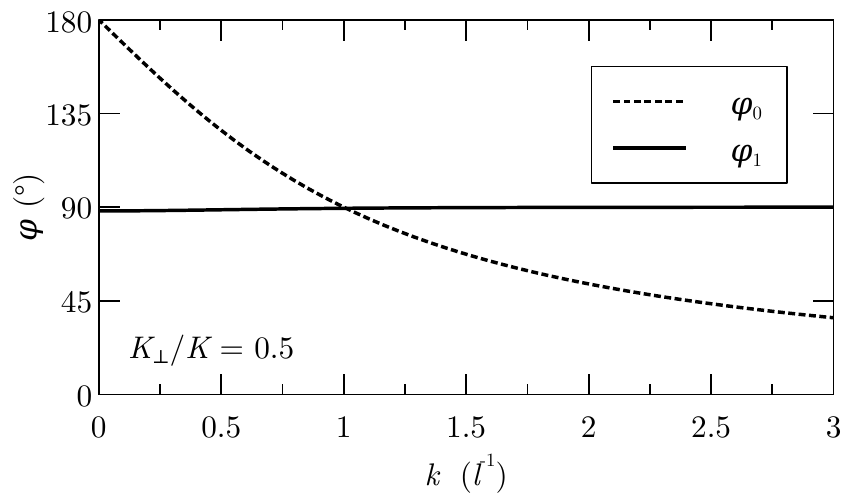}
  \caption{\label{fig:kxdep}
    Dependence of the phase shifts $\varphi_0$ [given by \equaref{eq:phi0basic}] and $\varphi_1$ [given by \equaref{eq:phasediff}] on wavenumber $k$. 
    For a spin wave propagating in the potitive-$x$ direction, the total phase shift induced by the domain wall is the sum $\varphi_0 + \varphi_1$.
    Notice that $\varphi_1$ is almost constant in $k$.
    Since only the part $\varphi_1$ depends on domain-wall chirality ($\pm$), the difference in phase shift between the two chiralities -- in other words, the phase difference obtained on the right-hand side of the interferometer shown in Fig.~2(d) of the main text -- is also almost independent of $k$.
  }
\end{figure}
Unlike $\varphi_0$, the chirality-dependent part $\varphi_1$ depends only very weakly on wavenumber $k$; it is, in fact, almost constant in $k$, as shown in \figref{fig:kxdep}.
We find numerically that this conclusion even holds if $D$ is not small or if the full dipolar interaction~\eqref{eq:dip1D} is taken into account in the scattering problem, at least in the regime where the wavelength is comparable to the domain-wall width or shorter.
This justifies our approach of taking the $k_x\rightarrow\infty$ limit in the calculation of $\varphi_\text{rel}$ in \secref{sec:phWKB} (the geometric part $\varphi_\text{geom}$ is independent of $k_x$ by definition).



\begin{thebibliography}{33}
\bibitem [{cap()}]{capnote1}%
  \BibitemOpen
  \href@noop {} {}\bibinfo {note} {Notice that the DMI strength $D$ used in
  Fig.~\ref{fig:sims}(b) is significantly lower than the value suggested by the
  plot of Fig.~\ref{fig:plotsmain}(b) for $\vartheta \approx 45^\circ$. The
  data in Fig.~\ref{fig:plotsmain} assume that the waveguide is infinite in the
  $y$ direction.}\BibitemShut {Stop}%
\end{thebibliography}
\end{document}